\documentclass[%
 reprint,
superscriptaddress,
showpacs,preprintnumbers,
 amsmath,amssymb,
 aps,
]{revtex4-1}

\usepackage{graphicx}
\usepackage{dcolumn}
\usepackage{bm}

\newcommand\redout{\bgroup\markoverwith{\textcolor{red}{\rule[.5ex]{2pt}{0.4pt}}}\ULon}

\usepackage[colorlinks=true,citecolor=blue]{hyperref}
\hypersetup{colorlinks=true,citecolor=blue,linkcolor=red,urlcolor=blue}

\begin{document}
\title{Electron mobility of a two-dimensional electron gas\\ at the interface of SrTiO$_3$ and LaAlO$_3$}
\author{A. Faridi}
\affiliation{Department of Physics, Sharif University of Technology, Tehran 14588-89694, Iran}
\affiliation{School of Physics, Institute for Research in Fundamental Sciences (IPM), Tehran 19395-5531, Iran}

\author{Reza Asgari}
\email{asgari@ipm.ir}
\affiliation{School of Nano Science, Institute for Research in Fundamental Sciences (IPM), Tehran 19395-5531, Iran}
\affiliation{School of Physics, Institute for Research in Fundamental Sciences (IPM), Tehran 19395-5531, Iran}

\author{A. Langari}
\affiliation{Department of Physics, Sharif University of Technology, Tehran 14588-89694, Iran}
\affiliation{School of Physics, Institute for Research in Fundamental Sciences (IPM), Tehran 19395-5531, Iran}
\affiliation{Center of excellence in Complex Systems and Condensed Matter
(CSCM), Sharif
University of Technology, Tehran 14588-89694, Iran}
\date{\today}

\begin{abstract}
We calculate the mobility of a two-dimensional electron gas residing at the interface
of LaAlO$_{3}$/SrTiO$_{3}$  following a three band Boltzmann approach at low temperature,
where carrier-charged impurity scattering process is assumed to be dominant.
We explain the anisotropic characteristic of the dielectric function, which is
a consequence of elliptical bands close to Fermi surface.
The screening effect, which weakens the long-range Coulomb interaction of the electron-impurity
is considered within the random phase approximation. Working at carrier densities high enough
to neglect the spin-orbit induced splitting of the bands, we find that the mobility varies
inversely with the cubic power of the carrier density ($n_{2D}^{-3}$)
in good agreement with the experimental
results. We also investigate the role of variable dielectric
constant of SrTiO$_ {3} $, the multiband nature of the system and interband interactions in
exploring this result.
\end{abstract}
\pacs{68.65.-k,72.10.-d,73.40.-c,73.20.-r}
\maketitle

\section{\label{sec:level1}Introduction}

The discovery of two-dimensional electron gas (2DEG) at the interfaces between two perovskite
insulators represented by the formula of ABO$_3$ , mainly LaAlO$_3$ (LAO) and SrTiO$_{3}$ (STO)
has attracted significant attention~\cite{ohtomo}.
Electrons in complex oxides with partially occupied d-orbitals interact with each other and with
the lattice. This gives rise to a wide extraordinary electronic properties, including respectable
mobility exceeding 5$\times 10^4$ cm$^2$V$^{-1}$s$^{-1}$,  low temperature superconductivity~\cite{reyren},
colossal magnetoresistivity, ferromagnetism, multiferroicity and a modulation of Rashba spin-orbit
coupling over a large range \cite{caviglia}. Therefore, oxide interfaces are exciting owing to the
opportunities emerging from broken symmetry, interfacial exchange interactions and spatial
confinement ~\cite{hwang}. Most importantly, the nature of the ground state present in the
complex oxide interfaces depends on the grow parameters of the LaAlO$_3$ layer, its layer thickness,
and the configuration of the heterostructures with different capping layers on top of
LaAlO$_3$~\cite{brink man}.

The carriers in 2DEG are confined within several nanometer (nm) near the interface, between polar (LAO)
and non-polar (STO) band insulators, as a result of broken symmetry the profile of distribution
has a sensitive dependence on the carrier density~\cite{exp}. When the gate voltage and the charge
carrier density are low, the 2D transport of heterointerfaces may present an inhomogeneous  character
since the distribution of itinerant carriers is not perfectly uniform. For instance, the charge
mobility decreases versus temperature ~\cite{thief}
at temperatures larger than $10$ K, while at low temperatures ($T<10 K$) it increases
with temperature for $n_{2D}\sim 10^{13}$ cm$^{-2}$. Furthermore, the increment of conductance versus
temperature is observed for $n_{2D} \leq 2\times 10^{13}$ cm$^{-2}$ at low temperatures.

The charge carrier mobility in LAO/STO has been measured by many groups and they have attempted
to increase the mobility using many different methods, such as lower growth temperature~\cite{caviglia},
polar solvents and conducting force microscopy~\cite{xie}. Recently, Sanders et al,~\cite{sanders}
measured the charge carrier mobility of doping the LAO side of the LAO/STO interface, where they showed
that doping has little effect on electron transport and found that a power law fits to the mobility
with respect to the electron density ($n_{2D}$) which behaves like $n_{2D}^{-3}$. In addition,
they explained that
the mobility is steeply anti-correlated with carrier concentration, $n_{2D}$, at low temperatures.

To shed light on the electronic mobility of LAO/STO interface and its properties, we, in this paper,
carry out a three band Boltzmann equation at low temperature. For this purpose, we consider the simple
three band Hamiltonian with only anisotropic nearest neighbor hopping and furthermore,
the screening effect is included through random phase approximation (RPA). In this regard,
we explain the anisotropic characteristic of the dielectric function of 2DEG at the interface.
This anisotropy is resulted from the presence of elliptical bands and hence dependence of the
interaction on both  $|\vec{k}-\vec{k'}|$ and the angle between them,
where $\vec{k}$ and $\vec{k'}$ are the electron momentum of the corresponding bands .
This structure is similar to 2DEG
structures where the effective mass is anisotropic and gives rise to anisotropic
transport properties~\cite{gold}. We emphasize on taking into account the variability of
the dielectric constant of SrTiO$_3$ with carrier density using a phenomenological relation.
We show that the experimental findings are properly approved considering the dielectric constant
reduction with confinement field or carrier density and also a  reasonable assumption of variable
impurity density. The significance of interband interactions specially at a lower carrier density is
discussed and we demonstrate that the exclusion of these interactions results in even faster decrease
of the mobility with carrier density.

This paper is organized as follows. The theory and model Hamiltonian are presented in
Sec. \ref{sec:level1} where the three band Boltzmann equation is derived and we further introduce
the screening potential in the RPA approach. Section \ref{sec:result} is devoted to our analytic
and numerical results for the screening and charge mobility of LAO/STO interface, in the presence
of charged long-range scatterers. We summarize and conclude our main findings in section \ref{sec:sum}.

\section{Theory and Method}
\subsection{\label{sec:level1}Three Band Model}

Several ab-initio calculations have shown that the 2DEG at the interface of band insulators LaAlO$_{3}$ and SrTiO$_{3}$ reside mostly on Ti-t$_{2g}$ orbitals (d$_{xy}$, d$_{yz}$, d$_{xz}$) \cite{pentcheva, zoran}. The crystal field reduces the energy of these orbitals compared to e$_g$ orbitals (d$_{x^2-y^2}$, d$_{z^2}$) by 2eV. On the other hand, d$_{xy}$ orbital is lower in energy by $\Delta$ due to the confinement of the electron gas along the out-of-plane axis, $\hat{z}$. Hopping between two t$_{2g}$ orbitals of the Ti atoms on neighboring sites is done through the overlap with the p orbitals of the Oxygen atoms between them. The hopping terms which couple t$_{2g}$ orbitals of different symmetry are very small and therefore we can ignore them \cite{khalsa}. The spatial extension of the d$_{xy}$ (d$_{yz}$/ d$_{xz}$) orbital is larger in $\hat{x}(\hat{y}/\hat{x})$ and $\hat{y} (\hat{z}/\hat{z})$ with respect to $\hat{z} (\hat{x}/\hat{y})$ and therefore their wave function overlap is stronger along these directions. This leads to the formation of three bands at the interface, namely; an isotropic circular band d$_{xy}$ with the same light effective mass along the $x$ and $y$ directions and two anisotropic elliptic bands d$_{yz}$ (d$_{xz}$) with a light mass in the $y$ ($x$) direction and a heavy one along the $x$ ($y$) direction. Here we assume that the conduction electrons only occupy the lowest t$_{2g}$ orbitals (even in moderate electron densities, these lowest bands contain more than $75\%$ of the electron density~\cite{khalsa}). Accordingly, a minimal tight-binding Hamiltonian can be written as \cite{khalsa, nayak}:
\begin{equation}
\mathbf{H} = \left(
\begin{array}{ccc}
\frac{\hbar^2k^2}{2m_L} & 0 & 0 \\
0 & \frac{\hbar^2k_x^2}{2m_H}+\frac{\hbar^2k_y^2}{2m_L}+\Delta & 0 \\
0 & 0 & \frac{\hbar^2k_x^2}{2m_L}+\frac{\hbar^2k_y^2}{2m_H}+\Delta
\end{array} \right),
\end{equation}\label{eq:H1}
in the basis (d$_{xy}$, d$_{yz}$, d$_{xz}$). There are two other important terms contributing to the main Hamiltonian of the system; atomic spin-orbit interaction and Rashba interaction~\cite{nayak, khalsa2}. The latter is caused by inversion symmetry breaking along $\hat{z}$ at the interface and permits hopping between adjacent d orbitals with different parity along the $\hat{z}$. The combination of these two additional interactions leads to a Rashba spin- orbit splitting and also orbital mixing of the bands which are highlighted near the band degeneracy points. Now if we choose the carrier density of the system large enough to stay far from these points, then we can neglect the Rashba spin-orbit term and the simple Hamiltonian of Eq.~(1) would be adequate to describe the properties of the system.

\subsection{Boltzmann Approach}

Boltzmann equation is a powerful tool to study the transport properties of an electron gas. The formalism is based on treating electrons as wave packets whose mean position and momentum obey classical equations of motion and all band structure effects appear in velocity which is related to energy dispersion. The mean free path of electrons is assumed to be much larger than their wavelength so that we can consider them as point-like quasiparticles. In this approach, our aim is to find the nonequilibrium distribution function of charge carriers responding linearly to an external field. The relaxation-time vector is
used to express the nonequilibrium part of the distribution function, which enters the Boltzmann equation~\cite{ziman}. Starting with the Boltzmann equation in 2D and generalizing it for a multiband system we thus have
\begin{equation}
\begin{split}
&-e\vec{E}.\vec{\upsilon}_n(\vec{k})(-\partial_\epsilon f_n^0)=\\
&\sum_{n'}\int\frac{d^2k'}{(2\pi)^2} W_{nn'}(\vec{k},\vec{k'})[f_n (\vec{k},\vec{E})-f_{n'} (\vec{k'},\vec{E})],
\end{split}
\end{equation}
where $e$ is the carrier charge, $\vec{E}$ is the electric field, $\vec{\upsilon}_n$ is the carrier velocity in band $n$ which is related to the energy dispersion $\epsilon_{\vec{k},n}$ through $\vec{\upsilon}_n=(1/\hbar)\nabla_{k} \epsilon_{\vec{k},n}$, $f_n (\vec{k},\vec{E})$ is the distribution function in band $n$ and $f_n^0$ is the same quantity at equilibrium. $W_{nn'}(\vec{k},\vec{k'})$ is the scattering rate from state $k$ in band $n$ to final state $k'$ in band $n'$. Since  we only assume  elastic scattering $W_{nn'}(\vec{k},\vec{k'})\propto\delta(\epsilon_{\vec{k},n}-\epsilon_{\vec{k'},n'})$ and microreversibility condition implies that $W_{nn'}(\vec{k},\vec{k'})=W_{n'n}(\vec{k'},\vec{k})$. For spherically symmetrical scattering rates and isotropic bands, the well known relaxation time approximation would be the exact solution of the above equation. But when we have an anisotropic system, this approximation can not capture all aspects of the transport phenomena. To account for anisotropies of the bands we follow an exact integral equation approach \cite{kovalev}.
Writing $\vec{E}$ and $\vec{k}$ as $\vec{E}=E(\cos\theta,\sin\theta)$ and $\vec{k}=k(\cos\phi,\sin\phi)$ and expanding the distribution function in Taylor series up to linear order in $\vec{E}$ we get
\begin{equation}
f_n(\theta,\phi)-f_n^0=E[A_n(\phi)\cos\theta+B_n(\phi)\sin\theta],
\end{equation}
where $A_n(\phi)=\partial_{E_x} f_n$ and $B_n(\phi)=\partial_{E_y} f_n$. Now substituting Eq. $(3)$ in $(2)$, we will eventually end up with two independent sets of integral equations
\begin{equation}
\begin{split}
\cos(\xi_n(\phi))&=\bar{W}_n(\phi)a_n(\phi)\\
&-\sum_{n'}\int d\phi'\bigg[\frac{v_{n'}(\phi')}{v_n(\phi)}W_{nn'}(\phi,\phi') a_{n'}(\phi')\bigg],
\end{split}
\end{equation}
\begin{equation}
\begin{split}
\sin(\xi_n(\phi))&=\bar{W}_n(\phi)b_n(\phi)\\
&-\sum_{n'}\int d\phi'\bigg[\frac{v_{n'}(\phi')}{v_n(\phi)}W_{nn'}(\phi,\phi') b_{n'}(\phi')\bigg],
\end{split}
\end{equation}
where $W_{nn'}(\phi,\phi')=(2\pi)^{-2}\int k'dk'W_{nn'}(k,k')$ and $\bar{W}_n(\phi)=\sum_{n'}\int d\phi'W_{nn'}(\phi,\phi')$. We also define $A_n(\phi)=-e\upsilon_n(\phi)(\partial_{\epsilon} f_n^0)a_n(\phi)$ and $B_n(\phi)=-e\upsilon_n(\phi)(\partial_{\epsilon} f_n^0)b_n(\phi)$.
In anisotropic systems the velocity is not parallel to $\vec{k}$ any more and it's magnitude is not constant in all directions. Therefore, we define the angle $\xi_n(\phi)$ as the angle between the velocity vector in band $n$ and $\hat{x}$ direction. In other words, we have $\vec{\upsilon}_n=v_n(\phi)(\cos\xi_n(\phi),\sin\xi_n(\phi))$. Eq. $(4)$ and $(5)$ are two independent systems of Fredholm equations of the second kind and can be solved numerically provided that we can find an analytical expression for both $v_n(\phi)$ and $\xi_n(\phi)$ for each band.
For the elliptical dispersion relation we can find such analytical relations as
\begin{eqnarray}
 v_n(\phi)&&=\frac{\hbar k_n(\phi)}{m_{(x,n)}m_{(y,n)}} \times \nonumber \\
 &&\sqrt{(m_{(x,n)})^2\sin^2(\phi)+(m_{(y,n)})^2\cos^2(\phi)},
\end{eqnarray}
\begin{equation}
\xi_n(\phi)=\arctan{\frac{m_{(x,n)}\sin(\phi)}{m_{(y,n)}\cos(\phi)}},
\end{equation}
where $k_n(\phi)$ is the angular dependence of momentum vectors of the elliptical bands at Fermi surface
 \begin{equation}
 k_n(\phi)=\frac{\sqrt{2\epsilon_{\rm F}}}{\hbar} \sqrt{\frac{m_{(x,n)}m_{(y,n)}}{m_{(x,n)}\sin^2(\phi)+m_{(y,n)}\cos^2(\phi)}},
  \end{equation}
with $m_{(x,2)}=m_{(y,3)}=m_H$ and $m_{(x,3)}=m_{(y,2)}=m_L$. It can be seen that for a circular band where $m_{(x,1)}=m_{(y,1)}=m_L$, we have $\xi_1(\phi)=\phi$, $\hbar k_1(\phi)={\sqrt{2\epsilon_{\rm F}}}$ and $v_1(\phi)$ does not depend on direction any more.
The scattering rate is evaluated using Fermi golden rule. Within the lowest order of Born approximation we have
\begin{equation}
W_{nn'}(\vec{k},\vec{k'})=\frac{2\pi}{\hbar}n_i|\langle n'\vec{k'}\mid\hat{V}\mid n\vec{k}\rangle|^2\delta(\epsilon_{\vec{k},n}-\epsilon_{\vec{k'},n'}),
\end{equation}
where $\hat{V}$ is the operator describing scattering and $n_i$ is the areal density of randomly distributed scatterers. Finally, if we can write $\mid\langle n'\vec{k'}\mid\hat{V}\mid n\vec{k}\rangle\mid^2$ as a function of $\phi$ and $\phi'$, then for elliptical bands $W_{nn'}(\phi,\phi')$ would be
\begin{equation}
 \begin{split}
&W_{nn'}(\phi,\phi')=(2\pi)^{-2}\int k'dk'W_{nn'}(k,k')\\
                  &=\frac{1}{2\pi \hbar^3}n_i|\langle n'\vec{k'}\mid\hat{V}\mid n\vec{k}\rangle|^2\big(\frac{\cos^2(\phi')}{m_{(x,{n'})}}+\frac{\sin^2(\phi')}{m_{(y,{n'})}}\big)^{-1}.
 \end{split}
\end{equation}

\subsection{Screening and Transport}

To understand the nature of scattering matrix elements $|\langle n'\vec{k'}\mid\hat{V}\mid n\vec{k}\rangle|$, which contain both interband and intraband terms, we should note that at low temperature the dominant scattering mechanism is charged impurity scattering. The strength of Coulomb interaction between electrons and charged impurity is modified by the screening of impurities by free electrons. In addition, many-body effects due to exchange and correlations, which can be described by a local-field correction~\cite{vignale}, have been neglected in our calculations due to the lack of expressions for anisotropic systems in the literature. We thus take into account the screening effect within RPA. So screened electron-impurity interaction would be~\cite{vignale}
\begin{equation}
\mathbf{V}(\vec{q})=(\mathbf{I}+\mathbf{V_b}(\vec{q})\boldsymbol{\chi}(\vec{q}))\mathbf{V_b}(\vec{q}),
\end{equation}
where $\mathbf{V}(\vec{q})$ is a $3\times3$ matrix expressing screened electron-impurity interaction in our three band model and $\mathbf{V_b}(\vec{q})$ is the matrix for unscreened electron-impurity interaction with the matrix elements of the form
\begin{equation}
[\mathbf{V_b}(\vec{q})]_{ij}= v(\vec{q})\exp^{-\mid \vec{q}\mid d_{ij}}=\frac{2\pi e^2}{\kappa\mid \vec{q}\mid}\exp^{-\mid \vec{q}\mid d_{ij}},
\end{equation}
in which $\kappa$ is an average dielectric constant of the surrounding medium. Since d$_{xy}$ orbital is more confined to the interface, an effective distance $d$ is assumed between this orbital and other orbitals to capture this effect~\cite{tolsma}, so that $d_{ij}$ is zero except for $(ij)=(12),(21),(13),(31)$. The RPA density-response matrix, $\boldsymbol{\chi}(\vec{q})$ is also defined as
\begin{equation}
\boldsymbol{\chi}(\vec{q})^{-1}=\boldsymbol{\chi^0}(\vec{q})^{-1}-\mathbf{V_b}(\vec{q}),
\end{equation}
here $\boldsymbol{\chi^0}(\vec{q})$ is the noninteracting density-response matrix with matrix elements $[\boldsymbol{\chi^0}(\vec{q})]_{ij}=\chi^0_i(\vec{q})\delta_{ij}$, where $\chi^0_1(\vec{q})$ is the density-response function of circular $d_{xy}$ band with bare-electron mass $m$ and $\delta_{ij}$ is the Koronecker delta~\cite{vignale}
\begin{equation}
 \chi^0_1(q;m) = \left\{
 \begin{array}{ll}
 -\frac{m}{\pi\hbar^2}\Big[1-\frac{1}{\tilde{q}}\sqrt{\tilde{q}^2-1}\Big] \qquad q \ge 2k_{\rm F},\\
 -\frac{m}{\pi\hbar^2} \qquad \qquad \qquad \qquad \;\;\; q<2k_{\rm F},\\
 \end{array} \right.
 \end{equation}
and $\tilde{q}={q}/{2k_{\rm F}}$. A connection between density-response function of an elliptical band with masses $m_x$ and $m_y$ and that of circular band is found by applying a Herring-Vogt transformation~\cite{herring} of the form $ k_x\to k_x'\sqrt{m_x/m_D}$ and $k_y\to k_y'\sqrt{m_y/m_D}$ with $m_D=\sqrt{m_xm_y}$. This rescaling enables us to transform the elliptical dispersion relation into a more convenient circular form $\epsilon(k')={\hbar^2k'^2}/{2m_D}$ and to define $k_{\rm F}$ for the elliptical band as $k_{\rm F}=\sqrt{2m_D\epsilon_F}/\hbar$. Therefore, following Ref. \cite{tolsma}, $\chi^0_2$ and $\chi^0_3$ (density-response functions of elliptical bands) can be found using the transformation below~\cite{tolsma}

\begin{eqnarray}
 \chi^0_{2}(\vec{q})=&&\chi^0_1(q';m_D)\mid_{|\vec{q'}|\to \big(q_x^2\sqrt{\alpha}+q_y^2\sqrt{\frac{1}{\alpha}}\big)^{1/2}},\nonumber\\
 \chi^0_{3}(\vec{q})=&&\chi^0_1(q';m_D)\mid_{|\vec{q'}|\to \big(q_x^2\sqrt{\frac{1}{\alpha}}+q_y^2\sqrt{\alpha}\big)^{1/2}},
\end{eqnarray}
where $m_D=\sqrt{m_H m_L}$ is equal for both bands and $\alpha=m_L/m_H$. Finally, these definitions will lead us to calculate matrix elements of screened electron-impurity interaction and then we can solve systems of Eq.  (4) and (5) to find nonequilibrium distribution functions of each band.

The current density in band $n$ of an electronic system is defined by
\begin{equation}
\vec{J}_n(\vec{E})=\int\frac{d^2k}{(2\pi)^2}e\vec{v}_n(\vec{k})f_n(\vec{k},\vec{E}).
\end{equation}
Substituting the calculated distribution function of Eq. (3) in the above expression and keeping in mind that $(-\frac{\partial f_n^0}{\partial\epsilon})\simeq\delta(\epsilon_{\vec{k},n}-\epsilon_{\rm F})$ (which for an anisotropic band depends on both $|k|$ and $\phi$ at low temperature), the conductivity of each band would be
\begin{equation}
 \begin{split}
&\sigma^n_{xx}/\sigma^n_{yy}=\frac{1}{(2\pi)^2}(\frac{2\epsilon_{\rm F}e^2}{\hbar^2}) \times\\
                                        &\int d\phi \{ \frac{(m_{(x,n)})^2\sin^2(\phi)+(m_{(y,n)})^2\cos^2(\phi)}{(m_{(x,n)}\sin^2(\phi))^2+m_{(y,n)}\cos^2(\phi)}\\
                                        &\times \cos(\xi_n(\phi)-\theta)[a_n(\phi)\cos\theta+b_n(\phi)\sin\theta]\},
\end{split} .
\end{equation}
with $\theta=0$ for $\sigma_{xx}$ and $\theta=\pi/2$ for $\sigma_{yy}$. We can also find the mobility of each band as $\mu_n={\sigma_n}/{n_{2D,n} e}$ with $n_{2D,n}$ being the electron density of band $n$. Finally, for a multiband system the total mobility is
\begin{equation}
\mu_T=\frac{\sum_n n_{2D,n}\mu_n}{\sum_n n_{2D,n}}.
\end{equation}
\begin{figure}
\centering
\includegraphics[width=.970\linewidth] {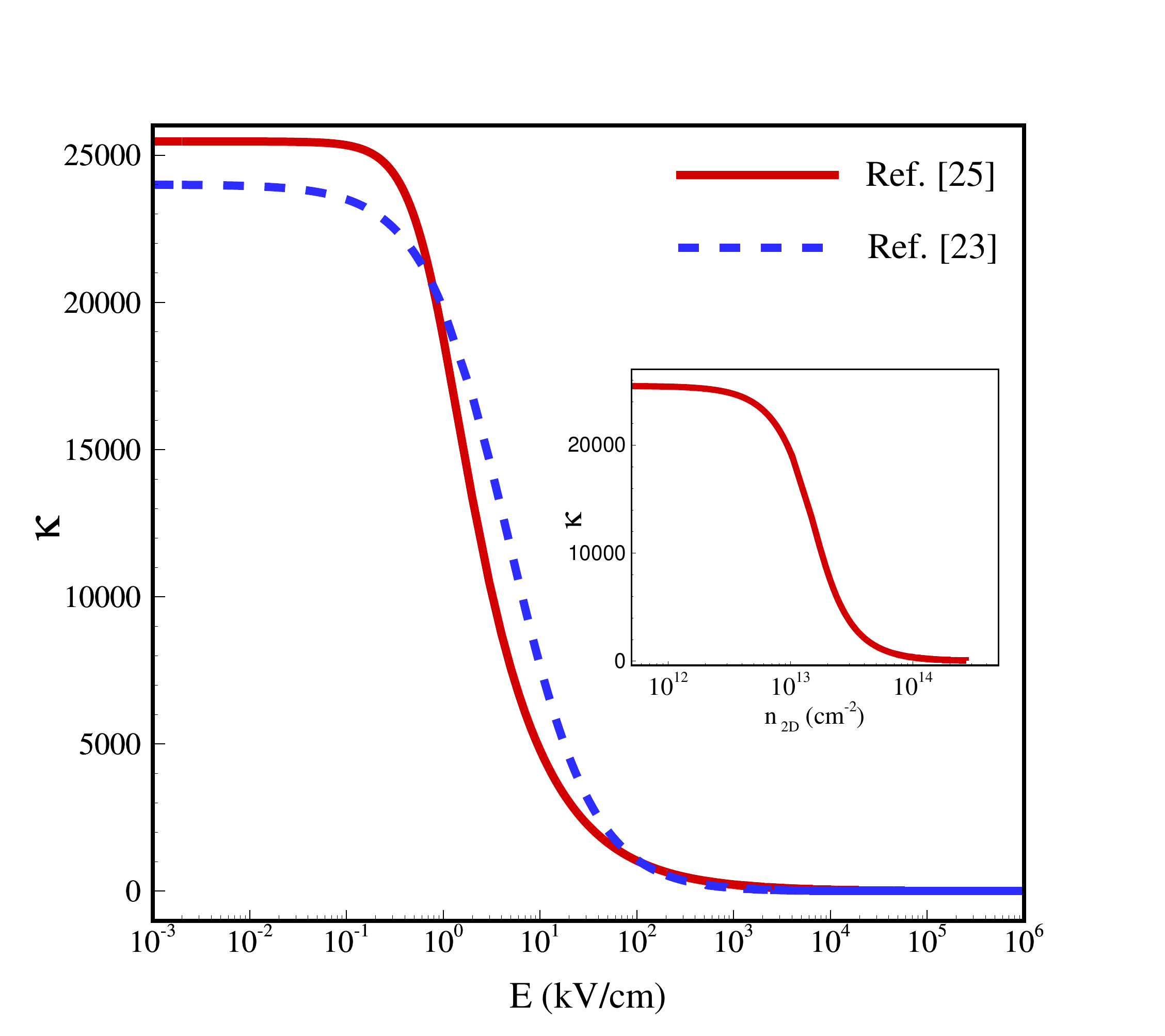}
\caption{\label{fig:epsart} (Color online) Electric field dependence of the dielectric constant of SrTiO$_3$. The solid line refers to Ref.~\cite{gariglio} and the dashed line is based on data from Ref. \cite{copie}. Inset: The dielectric constant as a function of the electron carrier density obtained from $\kappa E=e n_{2D}$.}
\end{figure}
In order to find transport properties of this system, dielectric constant of SrTiO$_3$ appearing in Coulomb interaction plays an important role. Although the low-temperature dielectric constant of SrTiO$_3$ is extremely large ($\kappa \sim 25000$), it also depends on the electric field at the interface. As the electron density and hence the electric field at the interface increases, the dielectric constant decreases nonlinearly. Fitting experimental data of Refs.~\cite{neville} and~\cite{ang}, Copie et al, proposed a field dependent dielectric constant of the form $\kappa=1+\frac{\chi_0}{1+E/E_C}$ with $\chi_0=24000$ and $E_C=4.7$  KV/cm~\cite{copie}. Recently, Gariglio et al, have also proposed a field dependent dielectric constant $\kappa=1+\frac{B}{[1+(E/E_0)^2]^{1/3}}$ based on their own experimental results together with estimations of Ref.~\cite{stengel} with $B=25462$ and $E_0=82213$ V/m~\cite{gariglio}. The displacement field at the interface is also related to the carrier density by $D=e n_{2D}$.  In Fig. 1, we compare the dielectric constant versus electric field for these two equations. As we can see there is a good agreement between the graphs specially for higher electric field (and higher electron density) region, which we are interested in. The  density dependence of the dielectric constant of the latter (which we have used in our calculations) is also shown in the inset of Fig. 1.

\section{Results and Discussions}\label{sec:result}
Numerical results for transport properties of the system are obtained in a range of carrier density $n_{2D}$ between $3\times 10^{13}$ cm$^{-2}$ and $4\times 10^{14}$ cm$^{-2}$. We set $m_L=0.68 m_e$ and $m_H=7.56 m_e$~  \cite{gariglio}. Band offsets between d$_{xy}$ and the other bands increases as the confinement field or electron density increases, but we assume a constant $\Delta=80$ meV in our calculations~\cite{tolsma}. For effective distance between d$_{xy}$ and other orbitals we choose $d=2a$~\cite{tolsma} where $a=3.9$ \AA, since its decrease with increasing the electron density does not affect final results effectively.

In order to find matrix elements of the screened electron-impurity interaction of Eq. (11), static dielectric function of the system within RPA approach is found as
 \begin{equation}
 \begin{split}
 \varepsilon(\vec{q})&=(1+v(q)\chi_1^0(\vec{q}))[1+v(q)(\chi_2^0(\vec{q})+\chi_3^0(\vec{q}))]\\
&-v^2(q)e^{-2qd}\chi_1^0(\vec{q})[\chi_2^0(\vec{q})+\chi_3^0(\vec{q})].
 \end{split}
 \end{equation}
Having calculated the dielectric constant of the system as a function of the electron density, we could calculate the screening of the system. In Fig. 2, we plot the dielectric function versus components of scattering momentum for $n_{2D}=8\times10^{13}$ cm$^{-2}$. The dielectric function of the system diverges when $\mathbf{|q|}\longrightarrow0$ and the anisotropy of the Fermi surface has resulted in a scattering angle dependent dielectric function.
\begin{figure}
\centering
\includegraphics[width=.970\linewidth] {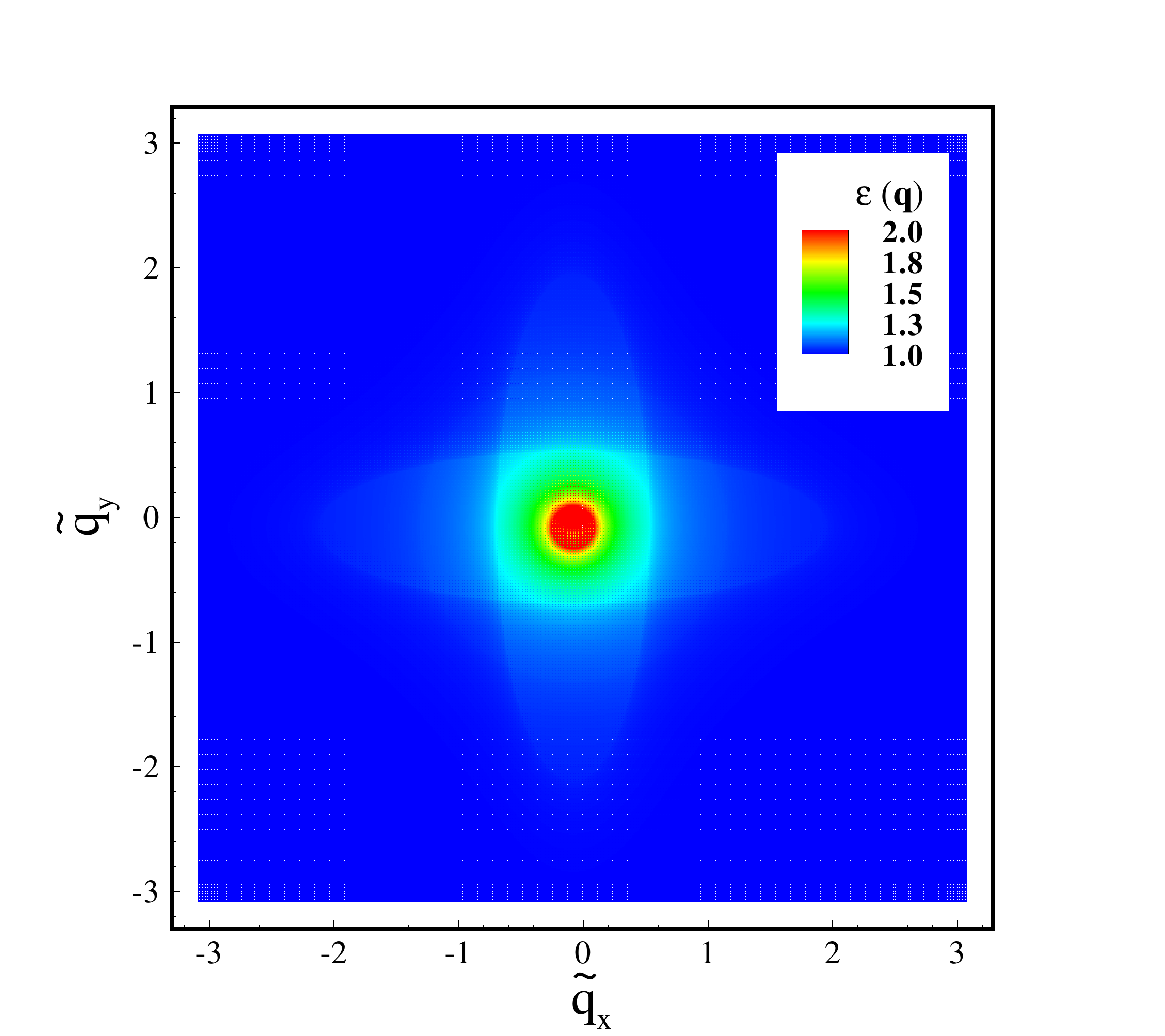}
\caption{\label{fig:epsart} (Color online) Dielectric function of the t$_{2g}$ interacting model versus components of the scaled momentum ($\tilde{\bf q}={\bf q}/\sqrt{2\pi n_{2D}}$). The electron density is $n_{2D}=8\times10^{13}$ cm$^{-2}$. The anisotropy of the Fermi surface has resulted in a scattering angle dependent dielectric function.}
\end{figure}

The solid line in Fig. 3 is the main result of this paper, where we show the mobility variations of the system with increasing carrier density. We find that the mobility of the system decreases sharply when the carrier density enhances and it can be scaled as $\mu \propto n_{2D}^{-3}$ in agreement with the avilable experimental data~\cite{wong, sanders}. This behavior is the direct consequence of considering the variations of dielectric constant of SrTiO$_3$ with the carrier density. Reducing the density of carriers results in an increase in dielectric constant, so that the screening effect would become more pronounced, which leads to electrons of higher mobility.
We also assume that the impurity density is not constant for the whole range of carrier density. This assumption is somehow physical, since whatever the mechanism of increasing the carrier density is, it would lead to an increase in charged impurity density as well, so that here we employ $n_i=0.01n_{2D}$. The dashed line in Fig. 3 shows the mobility of the system for a constant impurity concentration $n_i=10^{11}$ cm$^{-2}$. The graph is plotted in a broad range of carrier density, so the assumption of constant $n_i$ leads to $n_i\approx 3\times10^{-4} n_{2D}$ for higher densities and obviously such a low impurity-carrier density ratio can not capture the proper physics of the system. As can be seen from Fig. 3, although the mobility of the system is reduced in this case, but the reduction slows down at higher densities. We should mention that because of the high mobility in the LAO/STO interface, we expect that multiple scattering effects are not important for this system.
In the inset we show the mobility of the system, assuming a constant dielectric constant $\kappa=50$ for SrTiO$_3$. The intense reduction of the mobility by increasing the carrier density is strongly suppressed, owing to the constant dielectric constant assumption confirming the significant role of the variable dielectric constant for explaining the system appropriately. The reduction of mobility in this case is caused by assuming a variable impurity density. We should note that in the presence of a constant dielectric constant and impurity density, the mobility of the system increases with the carrier density, because the scattering centers do not change. But since the mobility is related to the impurity density as $\mu\propto 1/n_i$, then assuming $n_i=0.01n_{2D}$ can lead to reduction of the mobility with increasing the carrier density.

\begin{figure}
\centering
\includegraphics[width=.970\linewidth] {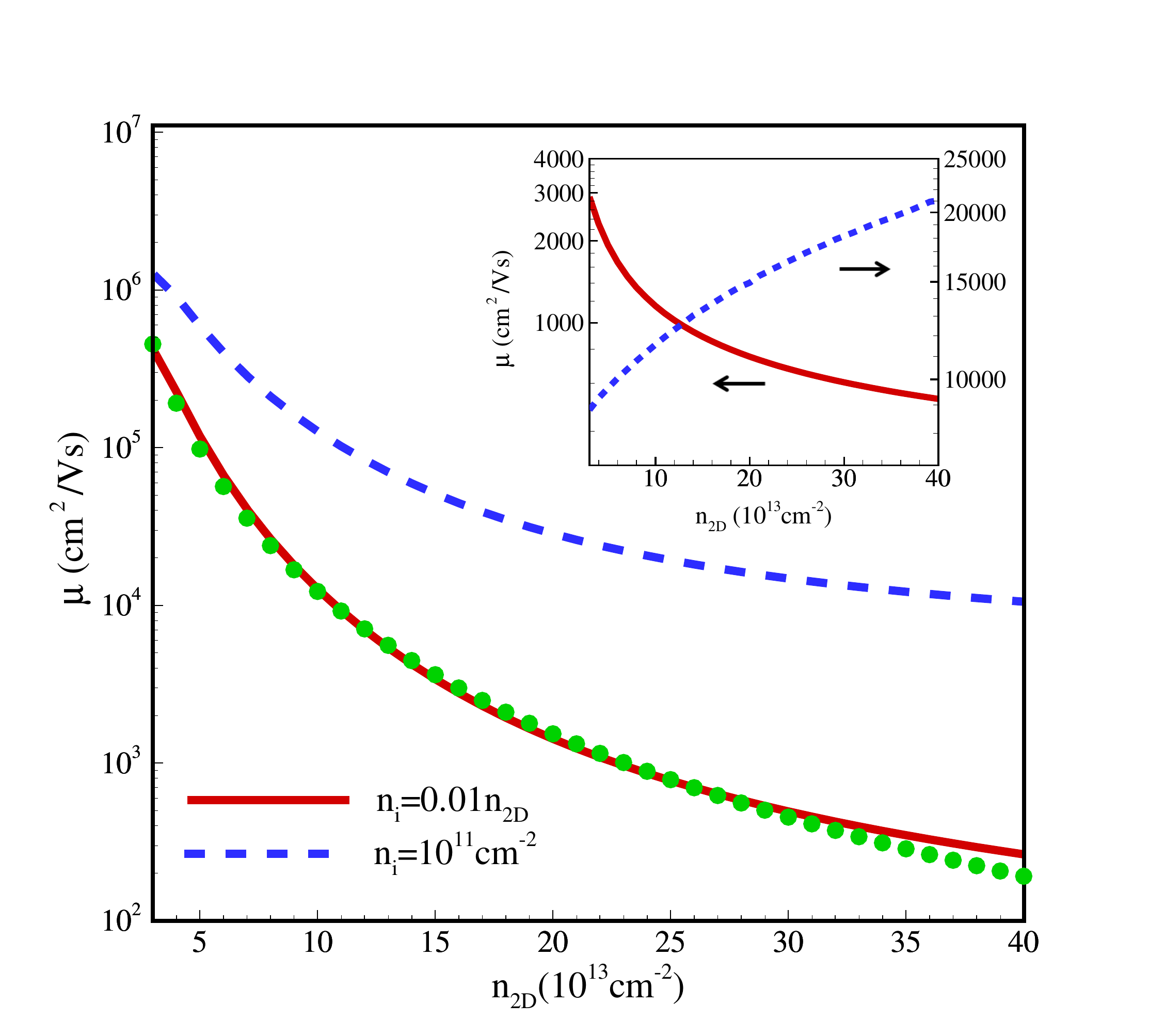}
\caption{\label{fig:epsart} (Color online) Electron mobility versus 2D carrier density considering $n_i=0.01 n_{2D}$ (solid line) and  $n_i=10^{11}$ cm$^{-2}$(dashed line). The mobility of the system decreases sharply when the carrier density enhances and the solid line is scaled as $\mu \propto n_{2D}^{-3}$. The symbols indicate a fit function. Inset: Mobility versus 2D carrier density considering a constant dielectric constant ($\kappa=50$) for the case of $n_i=0.01 n_{2D}$(solid line) and $n_i=10^{11}$ cm$^{-2}$(dashed line).}
\end{figure}

Taking into account the interband contributions in electron-electron interaction is also important to obtain  $\mu \propto n_{2D}^{-3}$ behavior of the mobility. In Fig. 4, we compare the mobility of the system, both in the presence (solid line) and in the absence (dashed line) of interband interactions. A faster decrease of the mobility with respect to the carrier density is found when the interband interactions are not considered. It can be seen that the effect of these interactions is crucial, especially in lower carrier densities where $|\vec{k}-\vec{k'}|$ is smaller for $\vec{k}$ and $\vec{k'}$ in different bands and therefore the interband interaction is stronger which results in considerable reduction of the mobility. As the carrier density increases, enhancement of $|\vec{k}-\vec{k'}|$ between different bands reduces the effect of interband interaction. It is also worthwhile to note that although the interband scattering processes lead to the enhancement of the mobility by a factor of 2, but it is not responsible for changing the behavior of the system effectively. In Fig. 4 we also show the mobility of an anisotropic 2DEG with $m^*=\sqrt{m_x m_y}$ and the same dielectric constant variation with carrier density as the main system. The mobility plotted in this case belongs to the direction with lower effective mass. Here again we can see roughly the same behavior as a result of the variable dielectric constant but the mobility is higher in this case with comparison to the t$_{2g}$ case, the direct result of multiband nature of the latter for which the existence of low mobility bands reduces the total mobility.
\begin{figure}
\includegraphics[width=.970\linewidth] {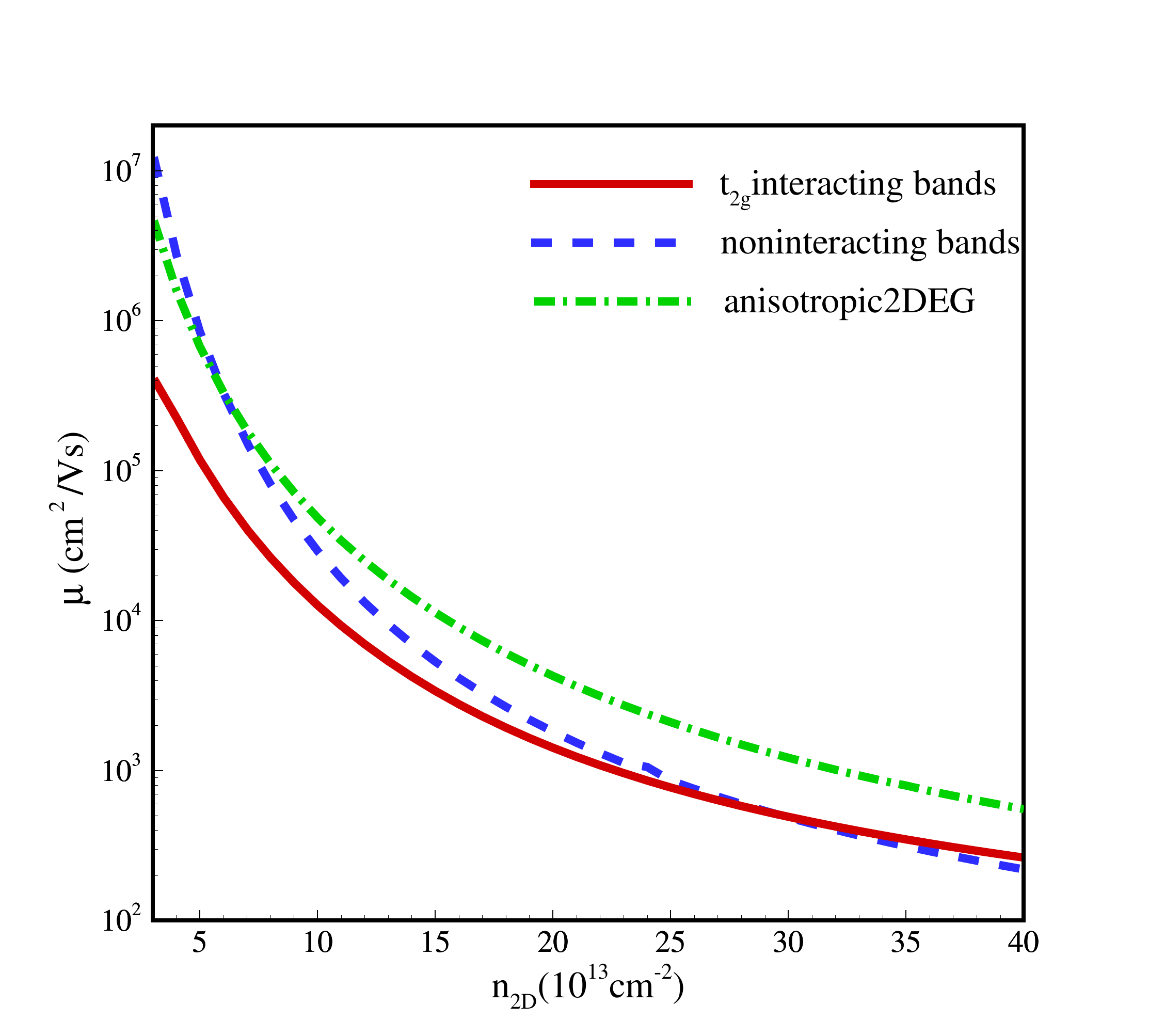}
\caption{\label{fig:epsart} {(Color online) Mobility versus 2D carrier density considering variations of dielectric constant as a function of electric field. Solid line shows the the result of calculations assuming both interband and intraband interactions. Dashed line is the mobility of the system ignoring the interband interactions and the dashed dotted line shows the mobility for an anisotropic 2DEG with the same effective mass of an ellipsoidal band and the same variable dielectric constant.}}
\end{figure}

Next, to investigate the contribution of bands in total mobility we plot the mobility of each band separately in Fig. 5. We show in the inset the density ratio of each band ${n_{2D,n}}/{n_{2D}}$, where $n=1$ $(d_xy)$ and 2 $(dyz,dxz)$, with respect to the total carrier density, $n_{2D}$. It follows from the graphs that in lower density region, the majority of carriers belong to d$_{xy}$ band with a relatively higher mobility. As the density of the system increases, d$_{xz}$ and d$_{yz}$ bands with an effective lower mobility become populated. Although these lower mobility bands reduce the total mobility of the system in higher density region, this effect is not strong enough to explain the peculiar reduction of the total mobility with density.
\begin{figure}
\includegraphics[width=.970\linewidth] {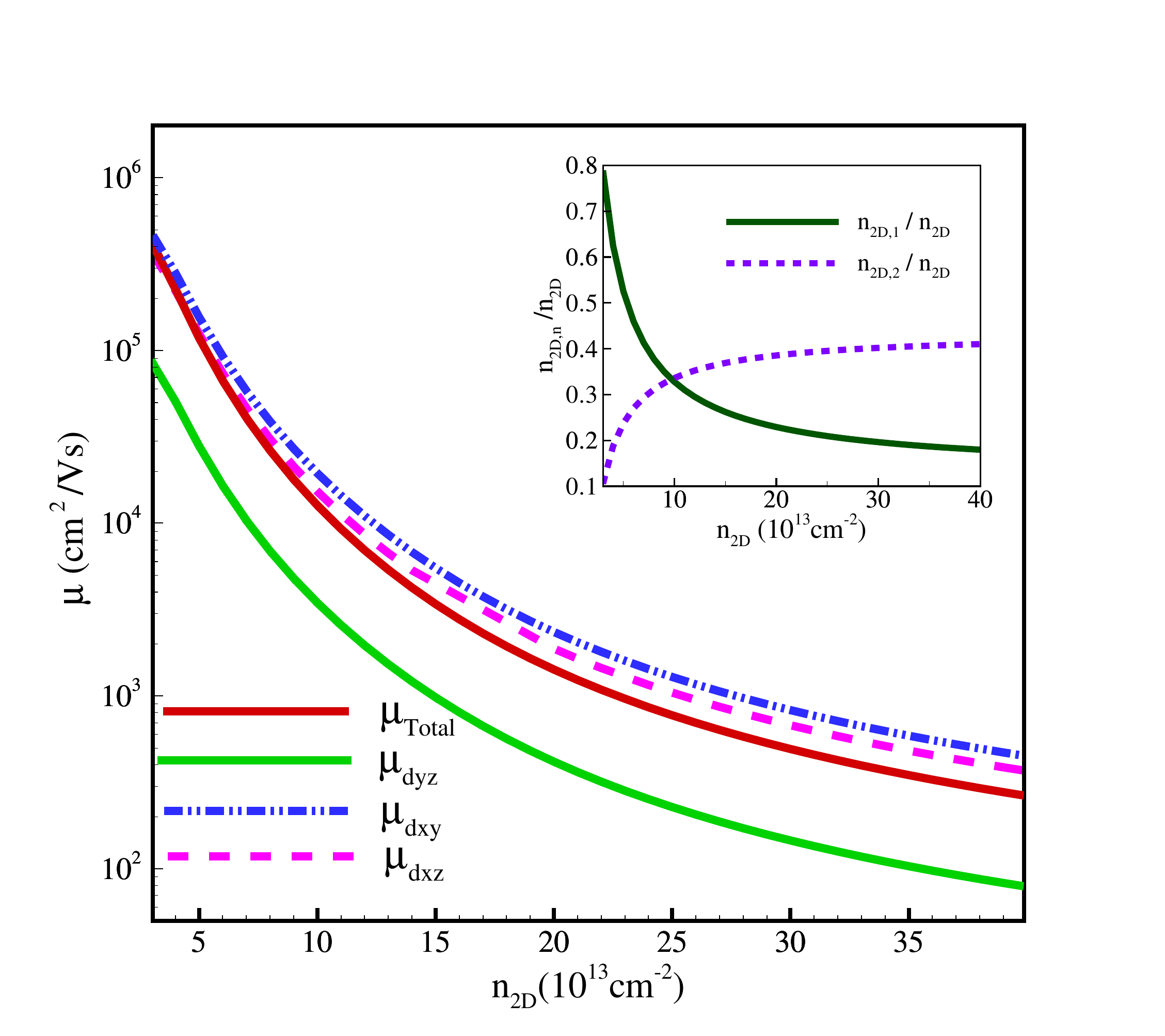}
\caption{\label{fig:epsart} {(Color online) Contribution of each band in the total mobility of the system. In lower density region, the majority of carriers belong to d$_{xy}$ band with a relatively higher mobility. As the density of the system increases, d$_{xz}$ and d$_{yz}$ bands with an effective lower mobility become populated. Inset: the density ratio of each band ${n_{2D,n}}/{n_{2D}}$, where $n=1$ and 2, with respect to the total carrier density.} }
\end{figure}
\section{Summary}\label{sec:sum}

In this paper, we have investigated transport properties of 2DEG at the interface of band insulators SrTiO$_3$ and LaAlO$_3$ pursuing an integral equation approach to solve a three band Boltzmann equation at low temperature. We performed our calculations in a range of carrier density where we could neglect spin-orbit interaction and employ the simple three band Hamiltonian with only anisotropic nearest neighbor hopping between {Ti}-{t$_{2g}$} orbitals. The scattering mechanism we employed is the electron- charged impurity scattering, which is dominant at low temperature. In order to obtain a more realistic picture of the scattering potential, screening effect was also included within the RPA approach. In this regard we have shown the anisotropic characteristic of the dielectric function of 2DEG at the interface resulting from presence of elliptical bands and hence dependence of the interaction on both  $|\vec{k}-\vec{k'}|$ and the angle between them.

To explain the peculiar reduction of the mobility of the system observed in the experiment we have emphasized on taking into account the variability of the dielectric constant of SrTiO$_3$ with carrier density using a phenomenological relation. We have shown that considering the dielectric constant reduction with confinement field or carrier density and also a  reasonable assumption of variable $n_i$ can properly approve the experimental findings. We should emphasize that the electron transport is sensitive to the origin of the dielectric constant, which its value changes by the 2D electron density. We have also discussed about the contribution of each band in total mobility and concluded that although at higher carrier densities, lower mobility bands become populated -which lowers the total mobility- but this effect can not explain the observed phenomenon by itself.

The significance of interband interactions specially at a lower carrier density was also discussed and we have demonstrated that the exclusion of these interactions results in even faster decrease of the mobility with carrier density. Interband scattering processes were indicated to be important as well to the extent that neglecting them, the mobility approximately increases by a factor of 2.

Although the overall behavior of the mobility is in agreement with experiment, the absolute value is overestimated in our calculations. This can be owing to the special choice of quantities such as effective mass and band offset. We should also note that even at low temperature there exist scattering centers other than that of charged impurity like interface roughness which can modify the total value of the mobility. On the other hand, it is believed that not all the electrons at the interface take part in conduction process,  but some of them are localized in trapped states at the interface \cite{zoran}. In this case they can alter the potential profile and the confinement field at the interface, but they do not contribute to the transport. Making use of this fact can reduce the value of the mobility of the system.

\section{Acknowledgments}
This work is partially supported by Iran Science Elites Federation under Grant No. 11/66332.

\end{document}